\documentclass[runningheads]{llncs}

\usepackage{graphicx}
\usepackage{amsmath}
\usepackage{amssymb}
\usepackage{array}
\usepackage{multirow}
\usepackage{booktabs}
\usepackage{xcolor}
\usepackage{hyperref}
\usepackage{makecell}
\usepackage{orcidlink}
\usepackage{pdfpages} 

\makeatletter
\newcommand{\printfnsymbol}[1]{
  \textsuperscript{\@fnsymbol{#1}}
}
\makeatother

\newif\ifshowrevisions
\showrevisionsfalse   
\newif\ifshowtodos
\showtodostrue        
\newif\ifshowrevlog

\ifshowrevisions
  \DeclareRobustCommand{\airevise}[1]{{\color{olive}#1}}
\else
  \DeclareRobustCommand{\airevise}[1]{#1}
\fi
\ifshowtodos
  \DeclareRobustCommand{\todo}[1]{{\color{orange}\bfseries [TODO:~#1]}}
  
\else
  \DeclareRobustCommand{\todo}[1]{}
  
\fi

\ifshowrevisions
  \DeclareRobustCommand{\fix}[1]{{\color{red}#1}}
\else
  \DeclareRobustCommand{\fix}[1]{#1}
\fi

\begin{document}
\pagestyle{empty}

\title{Directional Dual-Head Supervision for Real-Time Source-Free Polyp Segmentation}

\titlerunning{Directional Dual-Head Supervision for SF Polyp Segmentation}

\newif\ifsubmission
\submissionfalse

\ifsubmission
  \author{Anonymous Author(s)}
  \authorrunning{Anonymous Author(s)}
  \institute{Paper ID \textbf{410}\\ Affiliation withheld for double-blind review}
\else
    \author{
    Hai-Dang Nguyen\inst{1,2}\orcidlink{0009-0002-6475-6299}\thanks{These authors contributed equally to this work.}
    \and
    Nhat-Khoi Le\inst{1,2}\orcidlink{0009-0000-7143-4226}\printfnsymbol{1}
    \and
    Huu-Thuan Huynh\inst{1,2}\orcidlink{0009-0003-3984-3355}
    \and
    Minh-Triet Tran\inst{1,2}\orcidlink{0000-0003-3046-3041}
    \and
    Trung-Nghia Le\inst{1,2}\orcidlink{0000-0002-7363-2610}\thanks{Corresponding author.}
    }

    \authorrunning{H.-D. Nguyen et al.}

    \institute{$^1$ University of Science, Ho Chi Minh City, Vietnam\\
        $^2$ Vietnam National University, Ho Chi Minh City, Vietnam\\
    \email{\texttt{\{nhdang231, lnkhoi23\}@clc.fitus.edu.vn, hhthuan@hcmus.edu.vn, \{tmtriet,ltnghia\}@fit.hcmus.edu.vn}}}
\fi

\maketitle

\begin{abstract}
Source-free domain adaptation (SFDA) is critical for medical image segmentation as it adapts a model to an unlabeled target domain without accessing source data, thereby preserving data privacy. However, its effectiveness is often limited by confirmation bias: self-training models tend to miss small polyps, hallucinate foreground, and produce imprecise contours. We propose MedRT-SFSeg, a real-time SFDA framework \fix{built on a dual-head one-to-one/one-to-many YOLO segmentation architecture. Mask-DHF fuses precision- and recall-oriented predictions into stable pseudo-instances, while DURR uses the sign of pixel-level dual-head disagreement as a directional cue to expand or shrink masks, and additionally recovers missed instances and suppresses hallucinations.} Seg-MARD further regularizes feature representations through mask-guided, region-stratified sampling. Experiments on bidirectional adaptation between Kvasir-SEG and CVC-ClinicDB show that MedRT-SFSeg achieves 80.43\%/80.77\% Dice and 5.53/8.44 mm ASD, \fix{improving the non-adapted baseline by 7.7 and 9.8 Dice points while running at more than 62 FPS end-to-end. These results suggest that segmentation-aware dual-head supervision can improve pseudo-label quality without sacrificing real-time inference.} Project page: \url{https://danielnguyen-05.github.io/MedRT-SFSeg}.

\end{abstract}

\section{Introduction}
\label{sec:intro}

Automatic polyp segmentation is crucial for computer-assisted colonoscopy, yet models trained on one endoscopic dataset often degrade substantially when deployed on another because of differences in imaging devices, illumination, and acquisition protocols~\cite{CVCDB,jha2019kvasir}. Source-free domain adaptation (SFDA) offers an attractive solution by adapting a source-trained model to an unlabeled target domain without accessing the original source data~\cite{yu2023source}. This setting is particularly relevant to privacy-constrained medical deployment, where transferring source images across institutions may be undesirable or prohibited. However, adapting a segmentation model without target annotations remains challenging: self-training depends on model-generated pseudo-labels, and errors in these labels can be repeatedly reinforced during adaptation~\cite{chen2021source}.

A key limitation of existing uncertainty-based SFDA methods is that they primarily treat uncertainty as trust decision. Unreliable pixels are typically filtered, down-weighted, or denoised to construct more reliable pseudo-labels~\cite{chen2021source,wu2023upl}. In segmentation, however, uncertainty can also arise at ambiguous foreground--background boundaries, where it is particularly important to determine how the current mask should be corrected~\cite{yue2024boundary,bui2025up2d}. An uncertain boundary may indicate that the current mask is too conservative and should expand, or that it has included spurious foreground and should shrink. Simply discarding such pixels can therefore remove useful evidence for difficult instances, while retaining them can propagate incorrect supervision. We argue that effective SFDA segmentation should go beyond estimating whether a prediction is uncertain and also estimate how the prediction should change.

Recent real-time source-free object detection provides a useful opportunity to obtain such directional evidence~\cite{vcr2026real}. \fix{YOLO26}~\cite{yolo26} exposes complementary one-to-one (O2O) and one-to-many (O2M) prediction heads: the O2O head produces conservative, duplicate-free predictions, whereas the O2M head provides denser evidence that can recover missed instances. Existing dual-head adaptation methods exploit this complementarity for reliable pseudo-label construction and feature regularization~\cite{vcr2026real}. However, their box-level formulation does not capture the dense spatial correspondence and boundary sensitivity of segmentation. In particular, directly transferring box-level fusion to masks cannot determine whether disagreement should move a segmentation boundary inward or outward.

We therefore introduce \textbf{MedRT-SFSeg}, a real-time source-free adaptation framework that converts dual-head disagreement into segmentation-specific corrective supervision. \textbf{Mask-DHF} extends dual-head fusion from bounding boxes to masks by matching O2O anchors with O2M witnesses, estimating instance-level reliability, and fusing stable mask evidence. Building on these reliable predictions, \textbf{DURR} (Dual-Head Uncertainty Reliability Routing) preserves the sign of O2O/O2M disagreement \fix{as a heuristic cue} for whether the Student prediction should expand or shrink its boundary, while reliability weighting suppresses corrections unsupported by stable evidence. DURR additionally enables reliable O2M predictions to rescue instances missed by the O2O branch and suppresses foreground hallucinations where both heads indicate background. Finally, \textbf{Seg-MARD} adapts representation diversification to segmentation through mask-guided sampling of foreground and background regions. All these components operate only during source-free adaptation and are discarded afterward, leaving the adapted O2O branch unchanged for real-time inference.

We evaluated MedRT-SFSeg on bidirectional adaptation between Kvasir-SEG~\cite{jha2019kvasir} and CVC-ClinicDB~\cite{CVCDB}. The proposed framework consistently improves over the source-only model while retaining over 62 FPS end-to-end inference. Our analysis suggests that signed dual-head disagreement provides useful directional cues for boundary correction, highlighting the benefit of exploiting disagreement structure rather than using uncertainty solely for filtering.

Our contributions are as follows:
\begin{itemize}
    \item We introduce \textbf{Mask-DHF}, a segmentation-specific extension of dual-head fusion that \fix{matches O2O anchors with O2M witnesses and fuses their soft masks using mask-stability-aware instance reliability.}

    \item We propose \textbf{DURR}, which, unlike uncertainty-based filtering, \fix{uses the sign of reliability-weighted dual-head disagreement to guide directional boundary correction}, complemented by missed-instance rescue and hallucination suppression.

    \item We develop \textbf{Seg-MARD} \fix{for mask-guided feature regularization} and demonstrate that the complete framework achieves effective bidirectional source-free adaptation while preserving the real-time deployment efficiency of the underlying YOLO segmentation model.
\end{itemize}

\section{Related Work}
\label{sec:related}

\subsection{Source-Free Domain Adaptation in Medical Segmentation}
\label{subsec:sfda_medical}

Recent medical SFDA frameworks have primarily focused on three methodological streams: generative synthesis, prototype alignment, and uncertainty-guided self-training. Generative medical SFDA methods, such as SFDA-FSM~\cite{yang2022fourier} and HEAL~\cite{shi2025heal}, reconstruct source-like tissue distributions using Fourier style mining or diffusion denoising to reduce appearance-level discrepancies between source and target domains. Prototype-based approaches, including ProtoContra~\cite{yu2023source} and IAPC~\cite{cao2024towards}, extract source-derived class prototypes from pretrained models and align target representations through optimal transport and contrastive learning. Self-training paradigms, such as DPL~\cite{chen2021source} and UPL-SFDA~\cite{wu2023upl}, instead exploit pixel-wise prediction uncertainty to identify and suppress unreliable target pseudo-labels. \fix{Recent methods leverage Segment Anything Model (SAM) for pseudo-label generation or refinement,
including IPLC~\cite{Zha_IPLC_MICCAI2024}, IPLC+~\cite{zhang2025iplcplus},
SRPL-SFDA~\cite{liu2025srpl}, and DFG~\cite{huai2025leveraging}.
Mu et al.~\cite{mu2025illumination} further adapt SAM to low-light scenes.
Despite benefiting from SAM's shape priors, these methods incur the
computational cost of a large foundation model during adaptation or
deployment.} Despite their methodological diversity and effectiveness on offline medical segmentation benchmarks, these approaches have largely been evaluated in terms of adaptation accuracy, while deployment efficiency has received comparatively less explicit evaluation, particularly under real-time inference constraints.

\subsection{Pseudo-Label and Boundary Uncertainty in Polyp Segmentation}
\label{subsec:polyp_boundary}

Existing SFDA methods improve pseudo-label quality through uncertainty estimation, progressive adaptation, or explicit refinement. DPL~\cite{chen2021source} and UPL-SFDA~\cite{wu2023upl} suppress unreliable pseudo-labels using prediction uncertainty, while LFC~\cite{zhang2025robust}, IPLC~\cite{Zha_IPLC_MICCAI2024}, and UP2D~\cite{bui2025up2d} further stabilize or refine target supervision. In polyp segmentation, uncertainty is often concentrated near ambiguous foreground--background boundaries, motivating boundary-aware models~\cite{yue2024boundary,shao2024polyper}. However, most uncertainty-based approaches treat uncertainty mainly as a signal for filtering, weighting, or denoising. Such an unsigned quantity (e.g., entropy or prediction variance) does not indicate whether an erroneous boundary should expand or contract. We therefore seek to exploit prediction disagreement not only as an uncertainty measure, but also as a directional cue for boundary correction.

\subsection{Real-Time SFDA and Dual-Head Architecture Transfer}
\label{subsec:rtsfda}

Recent real-time source-free object detection (RT-SFOD) methods exploit the native O2O/O2M structure of YOLO architectures to construct reliable pseudo-labels through Dual-Head Fusion (DHF) and regularize representations through MARD~\cite{vcr2026real}. This architecture-aware paradigm enables source-free adaptation while retaining the efficient O2O head for deployment. However, these mechanisms were developed for bounding-box detection and do not directly address the dense spatial correspondence and boundary sensitivity required by medical segmentation. This motivates extending dual-head source-free adaptation from bounding boxes to pixel-level masks.

\section{Proposed Method}
\label{sec:method}

\subsection{Overview}
\label{subsec:overview}

Let $f_\theta$ be a YOLO-based segmentation model pretrained on labeled source data $\mathcal{D}_s=\{(x_i^s,y_i^s)\}$, which is unavailable during adaptation. Given only unlabeled target images
$\mathcal{D}_t=\{x_i^t\}$, our goal is to adapt $f_\theta$ to the target domain without accessing any source-domain images or annotations.

Specifically, we build upon the \textsc{YOLO26-S-Seg} architecture~\cite{yolo26}, which contains a duplicate-free one-to-one (O2O) head and a dense one-to-many (O2M) head. We exploit their complementary behavior as an intrinsic source of supervision under domain shift: the O2O head provides conservative, precision-oriented predictions, while the O2M head offers denser evidence for recovering missed instances and correcting boundaries.

For target-domain adaptation, we adopt a Mean-Teacher paradigm with a Student $f_{\theta_S}$ and an exponential moving average (EMA) Teacher $f_{\theta_T}$. Before self-training, Adaptive Batch Normalization (AdaBN)~\cite{adabn} recomputes the BatchNorm running statistics using unlabeled target images, producing target-initialized parameters $\theta_0$. Both networks are initialized from these parameters:
\begin{equation}
\theta_S^{(0)}=\theta_T^{(0)}=\theta_0.
\end{equation}
During adaptation, the Teacher parameters are updated from the Student as
\begin{equation}
\theta_T
\leftarrow
\mu\theta_T+(1-\mu)\theta_S,
\qquad \mu=0.999.
\label{eq:ema}
\end{equation}

\begin{figure}[t!]
\centering
\includegraphics[width=\textwidth]{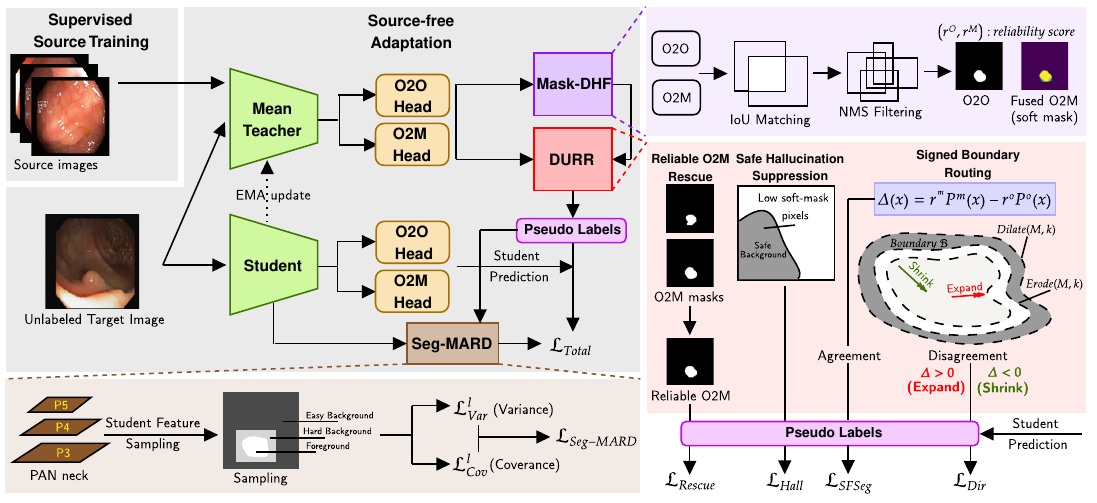}
\caption{Overview of \textsc{MedRT-SFSeg}. A source-trained dual-head segmentation model is first adapted to the target domain with AdaBN and Mean-Teacher self-training. Mask-DHF constructs reliable pseudo-instances from O2O/O2M predictions, while DURR performs reliability-aware signed routing and Seg-MARD regularizes multi-scale features. Only the adapted O2O Student is used at inference.}
\label{fig:pipeline}
\end{figure}

The overall adaptation process is illustrated in Fig.~\ref{fig:pipeline}.
The proposed components are described below, starting with Mask-DHF for
reliable pseudo-instance construction, followed by DURR for directional
dual-head routing and Seg-MARD for representation regularization.

\subsection{Mask-DHF: Dual-Head Pseudo-Instance Construction}
\label{subsec:maskdhf}

Mask-DHF extends the Dual-Head Fusion (DHF) strategy~\cite{vcr2026real}
from bounding-box detection to pixel-level segmentation. The core objective
is to leverage the O2O prediction as a conservative instance anchor while
{using} O2M predictions as complementary evidence to recover missed
instances and refine mask geometry. For every {Teacher} prediction, we consider its
confidence score $s$, bounding box, and soft mask probability map $P$.

An O2O prediction is retained as an anchor if {$s_{\mathrm{O2O}} \geq 0.5$. O2M predictions are filtered with the same confidence threshold and a minimum mask area of 16 pixels.} Each valid O2M prediction
is then associated with the {best-overlapping} same-class O2O anchor {in terms of} box IoU. Pairs with
$\mathrm{IoU} \geq 0.5$ are categorized as matched witnesses, whereas {O2M predictions}
with $\mathrm{IoU} \leq 0.2$ {to every anchor} are treated as potential coverage candidates
for instances missed by the O2O branch. To reduce redundancy, coverage
candidates are further filtered using class-wise Non-Maximum Suppression
(NMS) with a threshold of $\tau_{\mathrm{dup}}=0.7$.

To quantify the reliability of an individual prediction, we measure its mask
stability $q^{\mathrm{stab}}$ {as} the IoU between two binary masks
{obtained} at different probability thresholds,
$\tau_{\mathrm{lo}}=0.4$ and $\tau_{\mathrm{hi}}=0.6$:
\begin{equation}
q^{\mathrm{stab}}
=
\mathrm{IoU}
\left(
\mathbf{1}[P \geq \tau_{\mathrm{lo}}],
\mathbf{1}[P \geq \tau_{\mathrm{hi}}]
\right).
\label{eq:mask_stability}
\end{equation}
The resulting instance-level reliability is defined as
\begin{equation}
r = \sqrt{s \cdot q^{\mathrm{stab}}},
\label{eq:instance_reliability}
\end{equation}
where $r$ is a scalar associated with the entire prediction rather than a
pixel-wise reliability map.

For an O2O anchor $i$, let $r_i^o$ denote its instance-level reliability.
Its matched O2M witnesses $\{j\}$ are fused into an aggregated soft mask
using reliability-aware weights:
\begin{equation}
P_i^m(x)
=
\sum_j \bar{w}_{ij} P_{ij}^m(x),
\qquad
\bar{w}_{ij}
=
\frac{r_{ij}^m}
{\sum_k r_{ik}^m},
\label{eq:mask_fusion}
\end{equation}
where $r_{ij}^m$ is the instance-level reliability of the $j$-th O2M
witness. The reliability of the aggregated O2M evidence is {the}
reliability-weighted average of its constituent witnesses:
\begin{equation}
r_i^m
=
\sum_j \bar{w}_{ij} r_{ij}^m.
\label{eq:o2m_reliability}
\end{equation}
Both $r_i^o$ and $r_i^m$ are scalar instance-level quantities.

Coverage candidates are accepted only if their instance-level reliability
exceeds a {target-calibrated} threshold
\begin{equation}
\tau_r = Q_{0.25}(R_T),
\label{eq:reliability_threshold}
\end{equation}
where {$Q_{0.25}$ is the first quartile and} $R_T$ denotes the O2M instance-reliability distribution collected from
the initial Teacher model on the target domain. This threshold is computed
{once,} independently for each adaptation direction.

\subsection{DURR: Dual-Head Uncertainty Reliability Routing}
\label{subsec:durr}

Mask-DHF determines which dual-head predictions can be trusted, but the two
heads may still disagree in regions where the O2O prediction is incomplete
or inaccurate. DURR exploits
this disagreement as a directional supervision signal. Instead of treating
disagreement only as an uncertainty magnitude, DURR preserves its sign to
{indicate} whether the Student should expand or shrink the O2O mask.

For an O2O anchor $i$ with soft map $P_i^o$ and instance-level reliability
$r_i^o$, and its aggregated O2M evidence $P_i^m$ with instance-level
reliability $r_i^m$, we define the signed, reliability-weighted disagreement
at each pixel location $x$ as
\begin{equation}
\Delta_i(x)
=
r_i^m P_i^m(x)
-
r_i^o P_i^o(x).
\label{eq:delta}
\end{equation}
Here, $r_i^o$ and $r_i^m$ are scalar instance-level reliabilities that
uniformly weight the corresponding pixel-wise soft masks{, so} $\Delta_i(x)$
remains a pixel-wise disagreement signal without {requiring a}
spatial reliability map.

\fix{We use the sign of $\Delta_i(x)$ as a heuristic directional cue} $\Delta_i(x)>0$ indicates that the reliability-weighted O2M evidence favors
foreground more strongly than the O2O anchor, suggesting expansion, whereas
$\Delta_i(x)<0$ indicates that the O2O foreground is not sufficiently
corroborated by O2M evidence, suggesting shrinkage. We only {use}
non-trivial routing signals satisfying
$|\Delta_i(x)| \geq 0.02$.

\emph{Mechanism 1: Reliable O2M Rescue}. Due to its precision-oriented
nature, the O2O head may fail to detect a genuine instance entirely. To
mitigate this, DURR ``rescues'' reliable O2M candidates that lack a
corresponding O2O anchor. An O2M prediction is instantiated as a
\emph{rescue anchor} if it satisfies high confidence
($s_m \geq \tau_{\mathrm{res}}^{\mathrm{conf}}$) and mask stability
($q_m^{\mathrm{stab}} \geq \tau_{\mathrm{res}}^{\mathrm{stab}}$).
Its instance-level reliability is
\begin{equation}
r_m = \sqrt{s_m q_m^{\mathrm{stab}}}.
\label{eq:rescue_reliability}
\end{equation}
The rescued O2M mask then provides foreground supervision through
\begin{equation}
\mathcal{L}_{\mathrm{rescue}}
=
\frac{
\sum_x W_R(x)
\cdot
\mathrm{BCE}\big(S(x),1\big)
}{
\sum_x W_R(x)+\epsilon
},
\qquad
\fix{W_R(x)=r_m P_m(x),}
\label{eq:lrescue}
\end{equation}
where \fix{$S(x)$ is the Student mask logit, $\mathrm{BCE}(\cdot,\cdot)$ is the binary cross-entropy applied to logits, and the spatial weight $W_R$ combine the rescued soft mask {$P_m$} with
its instance-level reliability. } {Rescue} supervision is {thus} spatially concentrated on pixels supported
by the O2M mask, while its overall strength is modulated by the reliability
of the rescued instance. This allows the Student to recover target-domain
instances that are missed entirely by the O2O branch.

\emph{Mechanism 2: Signed Boundary Routing}. This mechanism targets contour refinement. {Rather than restricting correction to the mask interior}, DURR {uses} a symmetric, two-sided boundary band
\begin{equation}
\mathcal{B}_i = \mathrm{Dilate}(M_i, k) \setminus \mathrm{Erode}(M_i, k), \quad k = 5,
\label{eq:band}
\end{equation}
\fix{where $k$ is the structuring-element size in pixels}. This band allows for both outward correction (under-segmentation) and inward correction (over-segmentation). Within $\mathcal{B}_i$, we define the Teacher's desired direction $d_T(x) = \mathrm{sign}(\Delta_i(x))$ and the Student's movement relative to the O2O anchor $d_S(x) = P_s(x) - P_i^o(x)${, where $P_s$ is the Student's soft mask}. The Student is optimized via a directional margin loss
\begin{equation}
\mathcal{L}_{\mathrm{dir}} = \frac{1}{|\mathcal{B}_i|} \sum_{x \in \mathcal{B}_i} w(x) \max\left(0, m - d_T(x)d_S(x)\right),
\label{eq:ldir}
\end{equation}
\fix{where $w(x) = |\Delta_i(x)| \cdot r_i^{\mathrm{pair}}$ weights the penalty according to the magnitude of the disagreement and the joint reliability of the two heads, with $r_i^{\mathrm{pair}} = \sqrt{r_i^o r_i^m}$ denoting the pairwise reliability of the O2O and O2M heads, and $m$ representing the margin.} The directional loss $\mathcal{L}_{\mathrm{dir}}$ is averaged over all anchors in the batch. This encourages the Student to make substantial probability shifts only when the disagreement is strong and supported by reliable evidence from both heads.

\emph{Mechanism 3: Safe Hallucination Suppression}. To prevent the model from generating ``hallucinated'' foreground in regions where neither head provides evidence, we identify a safe background {region} $B_{\mathrm{safe}}=\{x: E_T(x)<\tau_{\mathrm{safe}}\}$\fix{, where $E_T(x)=\max\big(P_o^{\mathrm{raw}}(x),P_m^{\mathrm{raw}}(x)\big)$ is the pixel-wise maximum over all raw (pre-filtering) Teacher O2O and O2M mask probabilities.} {Student pixels predicted as confident foreground inside this region, $H=\{x\in B_{\mathrm{safe}}: P_s(x)>\tau_{\mathrm{student}}\}$, are penalized by}
\begin{equation}
\mathcal{L}_{\mathrm{hall}} = -\frac{1}{|H|} \sum_{x \in H} \log(1 - P_s(x)).
\label{eq:lhall}
\end{equation}
This confines background supervision strictly to areas where both native heads agree there is no foreground.

\subsection{Seg-MARD: Mask-Guided Representation Diversification}
\label{subsec:segmard}

Inspired by the representation diversification of MARD~\cite{vcr2026real}, Seg-MARD adapts feature-level regularization {to} segmentation by replacing coarse box-based sampling with pseudo-mask-guided regions. For each feature level $\ell \in \{P3, P4, P5\}$ {of the PAN neck}, we sample {feature} vectors from three zones: foreground ($M_i$), hard background (inside the bounding box but outside $M_i$), and easy background (outside all instances).

For each level, we apply a variance loss $\mathcal{L}_{\mathrm{var}}^\ell$ to prevent feature collapse and a covariance loss $\mathcal{L}_{\mathrm{cov}}^\ell$ to reduce inter-channel redundancy:
\begin{equation}
\mathcal{L}_{\mathrm{var}}^\ell = \frac{1}{C} \sum_c \max\left( 0, \gamma - \sqrt{\mathrm{Var}(Z_{\ell,c}) + \epsilon} \right), \quad \mathcal{L}_{\mathrm{cov}}^\ell = \frac{1}{C(C-1)} \sum_{p \neq q} \mathrm{Cov}(\tilde{Z}_{\ell})_{pq}^2,
\end{equation}
\fix{where $Z_\ell$ stacks the sampled $C$-dimensional vectors at level $\ell$ and $\tilde{Z}_\ell$ is its mean-centred version.} The total representation loss is $\mathcal{L}_{\mathrm{SegMARD}} = \sum_\ell (\alpha\mathcal{L}_{\mathrm{var}}^\ell + \beta\mathcal{L}_{\mathrm{cov}}^\ell)$, {with} $\gamma=1$, $\alpha=1$, {and} $\beta=0.1$. This {encourages} target-adapted features {to} remain diverse and discriminative across spatial scales.

\subsection{Overall Objective}
\label{subsec:objective}

The Student is optimized using a combined objective that stabilizes the self-training process while incorporating dual-head guidance:
\begin{equation}
\mathcal{L}_{\mathrm{total}} = \mathcal{L}_{\mathrm{SFSeg}} + \lambda_M(t)\mathcal{L}_{\mathrm{SegMARD}} + \lambda_D(t) \left( \lambda_{\mathrm{dir}}\mathcal{L}_{\mathrm{dir}} + \lambda_{\mathrm{res}}\mathcal{L}_{\mathrm{rescue}} + \lambda_{\mathrm{hall}}\mathcal{L}_{\mathrm{hall}} \right).
\label{eq:total}
\end{equation}
\fix{Here, $\mathcal{L}_{\mathrm{SFSeg}}$ is the standard YOLO26 segmentation loss}. The scheduled weights $\lambda_M(t)$ and $\lambda_D(t)$ gradually introduce regularization as the self-training process stabilizes.

\section{Experiments}
\label{sec:experiments}

\subsection{Implementation Details}
\label{subsec:implementation}

The adaptation stage (Mean-Teacher + Mask-DHF + DURR + Seg-MARD)
is optimized with SGD (momentum 0.937, Nesterov, weight decay
$5\times10^{-4}$), an initial learning rate of $1\times10^{-4}$
with cosine annealing to $0.01\times$ the initial rate,
gradient clipping at 10, batch size 4, and 60 epochs.
\fix{The Teacher is updated by EMA with $\mu=0.999$ at the end of each epoch.} All models are trained and evaluated identically in both adaptation directions on a single NVIDIA RTX 4060 Ti GPU.

\subsection{Datasets}
\label{subsec:setup}

We evaluate \textsc{MedRT-SFSeg} in both source-free adaptation directions{, Kvasir-SEG$\rightarrow$CVC-ClinicDB and CVC-ClinicDB$\rightarrow$Kvasir-SEG~\cite{jha2019kvasir,CVCDB}}, without changing any component of the method between directions; only the source checkpoint and the roles of $\mathcal{D}_s$ and $\mathcal{D}_t$ are swapped.

{Kvasir-SEG}~\cite{jha2019kvasir} contains 1{,}000 endoscopic polyp images with pixel-wise annotations; we use a {fixed random} split of 800 training / 200 validation images (seed 29) when Kvasir-SEG is the source domain, and the full image set as unlabeled target data when Kvasir-SEG is the target domain. {{CVC-ClinicDB}}~\cite{CVCDB} contains 612 frames extracted from colonoscopy video sequences with binary polyp masks; the full set is used analogously as source or unlabeled target data depending on {the} adaptation direction. Ground-truth masks for the target domain are used only for post-training evaluation, never for adaptation, in both directions. 

\subsection{Comparison with State-of-the-Art Methods}
\label{subsec:comparison}

In our experiments, \textbf{Baseline} denotes the source-only model trained on the labeled source domain and directly evaluated on the target domain without adaptation, while \textbf{Supervised} represents the target-supervised upper bound, where training and evaluation are both performed on the target domain. Table~\ref{tab:comparison} compares MedRT-SFSeg with representative SFDA methods in both adaptation directions. Dice and ASD are reported as mean$\pm$std across target images. \fix{As Kvasir-SEG and CVC-ClinicDB lack physical spacing metadata, we use unit spacing (1 pixel = 1 mm) at the evaluation resolution, following Shi Y. et al.~\cite{shi2025heal}.}

\begin{table}[t!]
\caption{Comparison with prior source-free domain adaptation methods in both adaptation directions between Kvasir-SEG and CVC-ClinicDB. Best in \textbf{\textcolor{red}{bold red}}, second-best \textcolor{blue}{\underline{underlined blue}}.}
\label{tab:comparison}
\centering
\resizebox{\textwidth}{!}{
    \footnotesize
    \setlength{\tabcolsep}{5pt}
    \begin{tabular}{l|c|c|c|c}
    \toprule
    \multirow{2}{*}{\textbf{Method}} & \multicolumn{2}{c|}{\textbf{Kvasir-SEG$\rightarrow$CVC-ClinicDB}} & \multicolumn{2}{c}{\textbf{CVC-ClinicDB$\rightarrow$Kvasir-SEG}} \\
    \cmidrule{2-5}
    & \textbf{Dice (\%) $\uparrow$} & \textbf{ASD (mm) $\downarrow$} & \textbf{Dice (\%) $\uparrow$} & \textbf{ASD (mm) $\downarrow$} \\
    \midrule
    Baseline & 72.8$\pm$33.0 & 5.9$\pm$12.5 & 70.9$\pm$25.2 & 9.8$\pm$18.5 \\
    Supervised & 96.0$\pm$4.3 & 0.6$\pm$0.6 & 90.3$\pm$16.6 & 3.7$\pm$10.2 \\
    \midrule
    ProtoContra~\cite{yu2023source} & 78.0$\pm$12.1 & \textbf{\textcolor{red}{0.9$\pm$0.6}} & \textcolor{blue}{\underline{68.9$\pm$26.4}} & 9.5$\pm$15.0 \\
    DPL~\cite{chen2021source} & 67.2$\pm$24.0 & 8.0$\pm$12.4 & 66.1$\pm$33.1 & \textbf{\textcolor{red}{7.8$\pm$17.7}} \\
    IAPC~\cite{cao2024towards} & 73.3$\pm$29.6 & 4.4$\pm$7.9 & 61.2$\pm$26.3 & 13.7$\pm$13.2 \\
    DFG~\cite{huai2025leveraging} & 19.1$\pm$14.7 & 97.5$\pm$17.6 & 26.8$\pm$18.1 & 87.5$\pm$19.5 \\
    UPL-SFDA~\cite{wu2023upl} & 67.1$\pm$14.6 & \textcolor{blue}{\underline{2.2$\pm$0.9}} & 60.2$\pm$31.3 & 11.4$\pm$17.2 \\
    HEAL~\cite{shi2025heal} & \textbf{\textcolor{red}{81.8$\pm$22.6}} & 4.3$\pm$8.2 & 66.5$\pm$28.8 & 9.2$\pm$11.6 \\
    IPLC~\cite{Zha_IPLC_MICCAI2024} & 34.8$\pm$11.7 & 34.7$\pm$22.9 & 45.3$\pm$21.2 & 19.2$\pm$8.4 \\
    SRPL~\cite{liu2025srpl} & 33.0$\pm$22.7 & 31.4$\pm$19.6 & 44.9$\pm$26.0 & 18.9$\pm$16.9 \\
    \midrule
    \textbf{MedRT-SFSeg (Ours)} & \textcolor{blue}{\underline{80.4$\pm$24.4}} & 5.5$\pm$10.7 & \textbf{\textcolor{red}{80.8$\pm$23.3}} & \textcolor{blue}{\underline{8.4$\pm$15.3}} \\
    \bottomrule
    \end{tabular}
}
\vspace{1mm}
\end{table}

The large gap between the baseline and the target-supervised upper bound confirms the substantial domain shift between the two datasets. MedRT-SFSeg consistently mitigates this degradation in both directions. On Kvasir-SEG$\rightarrow$CVC-ClinicDB, our method achieves 80.4\% Dice, ranking second among the compared SFDA approaches and remaining close to HEAL {(81.8\%)}. {In} the reverse direction, MedRT-SFSeg achieves the best Dice score of 80.8\%, improving the source-only baseline by {9.8} points and outperforming the strongest competing SFDA method by 11.9 points. Its ASD of 8.4mm is the second-best result in this direction, indicating that the gain in region overlap is achieved while maintaining competitive boundary accuracy. Overall, MedRT-SFSeg improves the source-only baseline by {7.7 and 9.8} Dice points in the two directions, respectively.

\subsection{Ablation Study}
\label{subsec:ablation}

\subsubsection{Effectiveness of Components}
\label{subsubsec:components}

Table~\ref{tab:ablation} evaluates the contribution of the main components of MedRT-SFSeg in both adaptation directions. Starting from the source-only baseline, Mask-DHF improves Dice by 3.08 points on K$\rightarrow$C and 4.10 points on C$\rightarrow$K. Adding Seg-MARD further improves Dice by 2.06 and 2.55 points, respectively, demonstrating the benefit of feature-level regularization. DURR provides larger gains over Mask-DHF, improving Dice by 3.19 and 4.29 points, respectively. Combining DURR and Seg-MARD yields the best performance, with further Dice gains of 1.41 and 1.44 points over Mask-DHF+DURR. \fix{The full model also reduces HD95 from 38.32 to 36.92 mm on K$\rightarrow$C and from 68.29 to 62.74 mm on C$\rightarrow$K relative to the baseline, indicating improved boundary quality.} These results suggest that output-space correction and feature-level regularization provide complementary benefits.

\begin{table}[t!]
\caption{Ablation study across both adaptation directions. Dice and IoU are
reported in \%, ASD and HD95 in mm. Best in \textbf{\textcolor{red}{bold red}}, second-best \textcolor{blue}{\underline{underlined blue}}.}
\label{tab:ablation}
\centering
\setlength{\aboverulesep}{0pt}
\setlength{\belowrulesep}{0pt}
\renewcommand{\arraystretch}{1.2}
\resizebox{\textwidth}{!}{
\begin{tabular}{l
    >{\centering\arraybackslash}p{1.35cm}
    >{\centering\arraybackslash}p{1.35cm}
    >{\centering\arraybackslash}p{1.35cm}
    >{\centering\arraybackslash}p{1.35cm} |
    >{\centering\arraybackslash}p{1.35cm}
    >{\centering\arraybackslash}p{1.35cm}
    >{\centering\arraybackslash}p{1.35cm}
    >{\centering\arraybackslash}p{1.35cm}}
\toprule
& \multicolumn{4}{c|}{\textbf{Kvasir-SEG $\rightarrow$ CVC-ClinicDB}}
& \multicolumn{4}{c}{\textbf{CVC-ClinicDB $\rightarrow$ Kvasir-SEG}} \\
\cmidrule(lr){2-5} \cmidrule(lr){6-9}
\textbf{Method}
& \textbf{Dice$\uparrow$} & \textbf{IoU$\uparrow$} & \textbf{ASD$\downarrow$} & \textbf{HD95$\downarrow$}
& \textbf{Dice$\uparrow$} & \textbf{IoU$\uparrow$} & \textbf{ASD$\downarrow$} & \textbf{HD95$\downarrow$} \\
\midrule
Baseline
& 72.75 & 65.54 & 5.92 & 38.32
& 70.94 & 63.07 & 9.81 & 68.29 \\
Mask-DHF
& 75.83 & 68.17 & 5.78 & 37.85
& 75.04 & 66.92 & 9.25 & 68.34 \\
Mask-DHF+Seg-MARD
& 77.89 & 70.02 & 5.62 & 37.79
& 77.59 & 69.21 & 8.84 & 68.07 \\
Mask-DHF+DURR
& \textcolor{blue}{\underline{79.02}}
& \textcolor{blue}{\underline{71.92}}
& \textcolor{blue}{\underline{5.58}}
& \textcolor{blue}{\underline{37.77}}
& \textcolor{blue}{\underline{79.33}}
& \textcolor{blue}{\underline{71.04}}
& \textcolor{blue}{\underline{8.59}}
& \textcolor{blue}{\underline{67.83}} \\
\textbf{Full MedRT-SFSeg}
& \textbf{\textcolor{red}{80.43}}
& \textbf{\textcolor{red}{72.57}}
& \textbf{\textcolor{red}{5.53}}
& \textbf{\textcolor{red}{36.92}}
& \textbf{\textcolor{red}{80.77}}
& \textbf{\textcolor{red}{72.52}}
& \textbf{\textcolor{red}{8.44}}
& \textbf{\textcolor{red}{62.74}} \\
\bottomrule
\end{tabular}
}
\end{table}

\subsubsection{Boundary Correction Analysis.}
\label{subsubsec:boundary_analysis}

\begin{figure}[t!]
\centering
\includegraphics[width=0.95\textwidth]{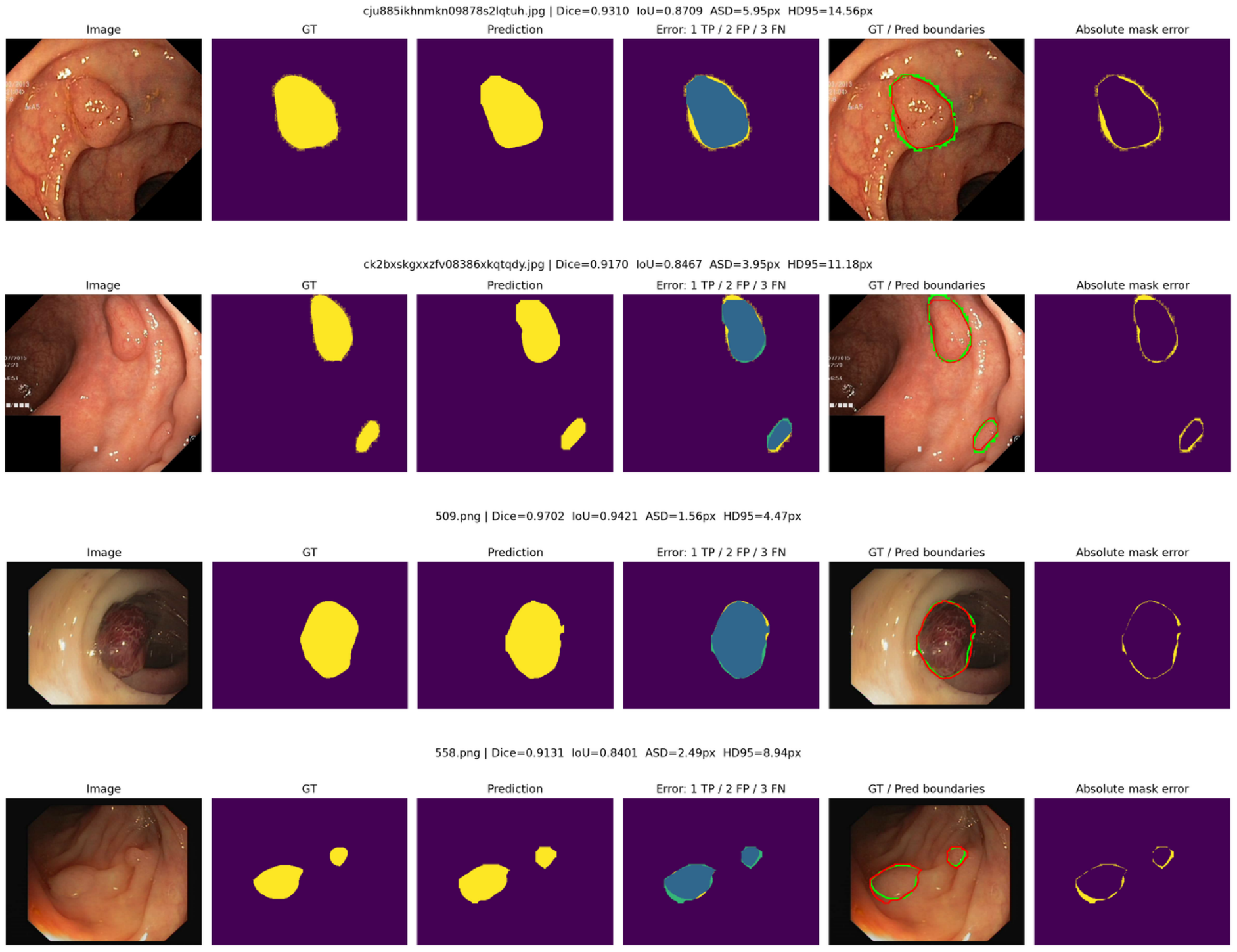}
\caption{Boundary correction analysis on representative target-domain cases. From left to right: input image; ground truth; Student prediction; error decomposition (true positive, false positive, and false negative regions); overlay of ground-truth (green) and predicted (red) boundaries; and absolute mask error. Residual errors are primarily localized near polyp contours, illustrating the boundary-sensitive behavior targeted by signed routing.}
\label{fig:qualitative}
\end{figure}

Figure~\ref{fig:qualitative} provides a qualitative analysis of the final Student predictions on representative target-domain cases. The error maps and boundary overlays show that residual errors are often concentrated around polyp contours rather than arising from gross false detections. This boundary-localized behavior is consistent with the motivation of DURR, which uses signed O2O/O2M disagreement to provide explicit expansion or contraction cues near uncertain
boundaries. The examples further illustrate that the adapted Student can recover both under-segmented and over-segmented regions while preserving the overall polyp structure. This supports the use of directional disagreement as a corrective signal rather than treating uncertainty solely as a criterion for filtering or suppressing pseudo-labels.

\subsection{Efficiency Evaluation}
\label{subsec:realtime}

Table~\ref{tab:realtime} reports the deployment-efficiency {and adaptation-cost} metrics {of} the final {MedRT-SFSeg model} in both adaptation directions, following the real-time protocol below.

\emph{Real-time protocol}. RTX 4060 Ti, $640\times640$ input, batch size 1, FP32, 100 warm-up iterations, 500 timed iterations, CUDA-synchronized; end-to-end (E2E) timing {includes pre- and post-processing but} excludes disk I/O.

\begin{table}[t!]
\caption{\fix{Deployment efficiency and adaptation cost} of the final MedRT-SFSeg model in both adaptation directions.}
\centering
\footnotesize {
\setlength{\tabcolsep}{6pt}
\begin{tabular}{l|r|r}
\hline
\textbf{Metric} & \textbf{Kvasir$\rightarrow$ClinicDB} & \textbf{ClinicDB$\rightarrow$Kvasir} \\
\hline
Full-graph params (M) & {11.43} & {11.43} \\
Fused deployment params (M) & {10.37} & {10.37} \\
Deployment GFLOPs & {34.05} & {34.05} \\
Checkpoint size (MB) & {22.32} & {22.34} \\
\hline
Network mean latency (ms) & {11.81} & {11.49} \\
Network FPS & 84.69 & 87.01 \\
E2E mean latency (ms) & {15.77} & {16.02} \\
E2E p95 latency (ms) & {15.83} & {16.11} \\
E2E FPS & 63.42 & 62.43 \\
\midrule
\textbf{Real-time} ($\geq$30 FPS) & \textbf{PASS} & \textbf{PASS} \\
\hline
\end{tabular}}
\vspace{1mm}
\label{tab:realtime}
\end{table}

The fused deployment parameter count (10.37M) matches the source model's native parameter count exactly and is identical across both directions, confirming that adaptation via AdaBN, Mean-Teacher self-training, Mask-DHF, DURR, and Seg-MARD adds no deployment-time overhead. End-to-end throughput exceeds the 30 FPS real-time threshold by more than $2\times$ in both directions, reaching 63.42 FPS and 62.43 FPS, respectively.

\subsection{Limitations}
\label{subsec:limitations}

A limitation of MedRT-SFSeg is its dependence on the dual-head O2O/O2M
architecture, which restricts direct transfer to conventional single-head
backbones. Moreover, the shared backbone may produce correlated errors under
severe domain shifts, reducing the effectiveness of reliability routing. Future
work will extend the framework to broader architectures and more diverse
multi-center clinical data.

\section{Conclusion}
\label{sec:conclusion}

We presented MedRT-SFSeg, a real-time source-free adaptation framework that exploits the spatial structure of dual-head disagreement for reliable target-domain learning. Mask-DHF constructs stable pseudo-instances from complementary O2O/O2M predictions, while DURR transforms reliability-weighted signed disagreement into directional boundary supervision, enabling boundary expansion and contraction, missed-instance recovery, and hallucination suppression. Seg-MARD further complements this process by regularizing multi-scale feature representations through mask-guided sampling. Across bidirectional adaptation between Kvasir-SEG and CVC-ClinicDB, MedRT-SFSeg achieves up to 80.8\% Dice while retaining over 62 FPS with a compact 10.37M-parameter deployment model. Overall, the results {suggest} that directional dual-head supervision \airevise{is} an effective and deployment-efficient strategy for source-free medical segmentation, improving target-domain accuracy without {adding} inference-time computation.

\begin{credits}
\ifsubmission
\else
\subsubsection{\ackname}
This research is funded by Vietnam National University - Ho Chi Minh City (VNU-HCM) under Grant Number A2025-18-04.
\fi
\end{credits}

\bibliographystyle{splncs04}
\bibliography{references}

\ifshowrevlog
\includepdf[pages=-]{revision_log.pdf}
\fi

\end{document}